\documentclass[a4paper,11pt]{article}

\pdfoutput=1 
\usepackage{jcappub}

\usepackage{graphicx}
\usepackage{longtable}
\usepackage{float}
\usepackage{dcolumn}
\usepackage{bm}
\usepackage{appendix}
\usepackage{multirow}
\usepackage{color}
\usepackage[utf8]{inputenc}
\usepackage{footmisc}
\usepackage[normalem]{ulem}

\newcommand{\mytilde}{\raise.17ex\hbox{$\scriptstyle\mathtt{\sim}$}}
\newcommand{\barr}{\begin{eqnarray}}
\newcommand{\earr}{\end{eqnarray}}
\newcommand{\bea}{\begin{eqnarray*}}
\newcommand{\eea}{\end{eqnarray*}}
\newcommand{\beq}{\begin{equation}}
\newcommand{\eeq}{\end{equation}}

\title {Can cosmic rotation resolve the Hubble tension? Constraints from CMB and large-scale structure}

\author[a,b]{Micol Benetti}\emailAdd{m.benetti@ssmeridionale.it}

\author[c]{David A. Cook}\emailAdd{dcook@ufba.br}

\author[c,d]{Saulo Carneiro}\emailAdd{saulocarneiro@on.br}

\affiliation[a]{Scuola Superiore Meridionale, Via Mezzocannone 4, I-80138 Napoli, Italy}

\affiliation[b]{Istituto Nazionale di Fisica Nucleare, sez. di Napoli, Via Cinthia 9, I-80126 Napoli, Italy}

\affiliation[c]{Instituto de F\'isica, Universidade Federal da Bahia, 40210-340, Salvador, BA, Brasil}

\affiliation[d]{Observat\'orio Nacional, 20921-400 Rio de Janeiro, RJ, Brasil}

\abstract{
We investigate a relativistic cosmological model with background rotation, sourced by a non-perfect fluid with anisotropic stress. A modified version of the CLASS Boltzmann code is employed to perform Monte Carlo Markov Chain analyses against Cosmic Microwave Background (CMB) and late-time datasets. The results show that current CMB data constrain the present-day rotation parameter to be negligible. As a consequence, the derived cosmological parameters remain consistent with the standard $\Lambda$CDM values. In contrast, late-time probes such as Type Ia supernovae (SNe) and Baryonic Acoustic Oscillations (BAO) allow for a higher level of rotation and yield an increased Hubble constant. However, this comes at the cost of a higher $\sigma_8$, which remains in tension with DES-Y3 measurement. Combining CMB, SNe and BAO data confirms the preference for non-rotation.}

\begin{document}
\maketitle

\section{Introduction}
\label{Introduction}

The persistent discrepancy between the value of the Hubble constant $H_0$ derived from {\color{black} Cepheid-calibrated SNe Ia} observations \cite{Riess:2019cxk,Riess:2021jrx} and that inferred from the Cosmic Microwave Background (CMB) \cite{Aghanim:2018eyx} -- the so-called Hubble tension -- remains one of the most significant unsolved problems in modern cosmology. Proposed solutions include potential systematic errors in the calibration of Cepheids \cite{cambridge} or type Ia supernovae, such as biases in redshift determination \cite{Tamara,EPJPlus} and reddening corrections \cite{polaco}. They also include a wide range of alternative cosmological models \cite{review}-\cite{Li:2019san}, which in general fail to reconcile the tension without disrupting the accurate fit of the full CMB anisotropy spectrum.

Among the alternative models, rotating cosmologies have a long history, beginning with G\"odel’s seminal solution to Einstein’s field equations \cite{Godel}. This model and its generalisations by Rebouças and Tiomno \cite{Tiomno}, and later by Korotkii and Obukhov \cite{Korotkii,Obukhov}, established the possibility of consistent cosmological solutions with global rotation. Importantly, these models showed that causality violations — a major concern in rotating spacetimes — can be avoided if the rotation parameter remains sufficiently small.
Recently, rotation has been revisited in the context of the Hubble tension by Szigeti et al. \cite{newton}, who investigated its effects within a non-relativistic framework. While suggestive, their approach lacked a fully relativistic treatment capable of addressing both the background dynamics and the detailed structure of cosmological perturbations.

In this work, we explore the implications of a relativistic rotating cosmological model with G\"odel-type metric originally proposed in 2002 \cite{saulo}, in which the rotation is sourced by a non-perfect fluid with anisotropic stress. The model introduces a small but finite vorticity in the cosmic expansion, offering a new mechanism to slightly modify the inferred value of $H_0$ while preserving the successes of the standard cosmological model.
Using the CLASS Boltzmann code, we implement the background dynamics of this rotating model to fit the observed CMB anisotropy angular power spectrum. 
{\color{black}
The analysis yields a precise determination of the cosmological parameters and shows that the inclusion of rotation does not lead 
to significant corrections in the Hubble constant.}
In this first study, only the background evolution is modified; perturbative corrections due to rotation are a higher order effect that will be deferred to future work.

The paper is structured as follows: in Section 2 we present the rotating model and its dynamical properties. Section 3 discusses the corresponding Hubble function and the determination of cosmological parameters. The numerical implementation and main results are shown in Section 4, followed by our conclusions in Section 5.

\section{Rotating spacetime}

The G\"odel type metric that we will consider is given by \cite{saulo}
\begin{equation} \label{1}
    ds^2 = a^2(\eta) \left[ \left( d\eta + l e^x dy \right)^2 - \left( dx^2 + e^{2x} dy^2 + dz^2 \right) \right],
\end{equation}
where $a$ is a scale factor, $l < 1$ is a positive parameter (the rotation parameter), $\eta$ is the conformal time and $x,
y, z$ are spatial coordinates. As we will see, for {\color{black} late times} the {\color{black} $l$-dependent terms} can be neglected in the Einstein equations, which means to consider, instead of metric (\ref{1}), the Bianchi type-III anisotropic metric
\begin{equation} \label{2}
    ds^2 = a^2(\eta) \left[ d\eta^2 - d\xi^2 - \sinh^2\xi\, d\phi^2 - dz^2 \right],
\end{equation}
written in cylindrical coordinates\footnote{For a study of inflationary perturbations in the context of the anisotropic metric (\ref{2}), see \cite{Thiago1,Thiago2}. For observational fittings with supernovae and the acoustic scale, see \cite{roberto,welber}.}. In spherical coordinates, it can also be written as
\begin{equation}
    ds^2 = a^2(\eta) \left[ d\eta^2 - d\chi^2 - \chi^2 d\theta^2 - \sinh^2 \left(\chi \sin \theta \right) d\varphi^2\right],
\end{equation}
which is reduced to a spatially flat {\color{black} Friedmann-Lema\^itre-Robertson-Walker (FLRW)} metric in the limit of small distances.

Metric (\ref{1}) respects the observed isotropy of cosmic microwave background (CMB) and does not cause observable parallax effects \cite{Mena}.\footnote{\color{black}{In this respect, it differs from other homogeneous metrics which imprint anisotropies on the CMB at the very background level, being in this way ruled out by observations \cite{thiago4}.}} Moreover, the closed
time-like curves characteristic of the G\"odel metric do not appear if $l<1$ \cite{Tiomno}. It describes an expanding and rotating universe, with an angular
velocity given, in comoving coordinates, by $\omega = l/2a$ \cite{Obukhov}. It is possible to see that, in the radiation-dominated epoch,
the parameter $l$ is constant,
while in the matter-dominated era it falls with $a$. Indeed, from conservation
of angular momentum we have $\epsilon \omega a^5 = $ constant, where $\epsilon$ is the energy density of the matter content. In the radiation epoch $\epsilon$ decreases with $a^4$, and so $\omega$ falls with $a$,
leading to a constant $l$. On the other hand, in the matter era $\epsilon$ decreases with $a^3$, so $\omega$
falls with $a^2$ and $l$ falls with $a$. 

Although it is not trivial, in general, to express global conservation laws in curved spacetimes, the above ansatz for the conservation of angular momentum leads to consistent continuity equations in the radiation and matter eras. Its derivative w.r.t. to time leads to
\begin{equation}
    \dot{\epsilon} + \epsilon \left( \frac{\dot{\omega}}{\omega} + 5 \frac{\dot{a}}{a} \right) = 0.
\end{equation}
Using $\omega = l/2a$ we obtain
\begin{equation}
    \dot{\epsilon} + \epsilon \left( \frac{\dot{l}}{l} + 4 \frac{\dot{a}}{a} \right) = 0.
\end{equation}
In the radiation era $p = \epsilon/3$ and $l$ is constant, while in the matter era we have $p = 0$ and $l a=$ constant. In both cases the above equation can be rewritten as the energy balance equation
\begin{equation}
    \dot{\epsilon} + 3 \frac{\dot{a}}{a} (\epsilon + p) = 0.
\end{equation}

From metric (\ref{1}) and considering $l$ as a function of time, the diagonal Einstein
equations are given by \cite{saulo}
\begin{equation} \label{epsilon}
    \epsilon a^4 = - \left( 1 - \frac{3l^2}{4} \right) a^2 + 3\left( 1-l^2 \right) \dot{a}^2 - 2 l \dot{l} a \dot{a},
\end{equation}
\begin{equation}
    p_1 a^4 = \left( \frac{l^2}{4} + \dot{l}^2 + l \ddot{l} \right) a^2 + \left( 1-l^2 \right) \dot{a}^2 - 2\left( 1-l^2 \right)a \ddot{a} + 4 l \dot{l} a \dot{a},
\end{equation}
\begin{equation}
    p_2 a^4 = \left( \frac{l^2}{4} \right) a^2 + \left( 1-l^2 \right) \dot{a}^2 - 2\left( 1-l^2 \right)a \ddot{a} + 2 l \dot{l} a \dot{a},
\end{equation}
\begin{equation}
    p_3 a^4 = \left( 1 - \frac{l^2}{4} + \dot{l}^2 + l \ddot{l} \right) a^2 + \left( 1-l^2 \right) \dot{a}^2 - 2\left( 1-l^2 \right)a \ddot{a} + 4 l \dot{l} a \dot{a},
\end{equation}
where the dot means a derivative with respect to the conformal time. 

The non-diagonal components of the Einstein equations give terms subdominant in the limits of small rotation or small scale factor. In fact, all of them are proportional to $l$ and its time derivative, and further scale with inverse powers of the scale factor. As in the matter era $l$ falls with $a$, those terms fall faster than the matter density and can be neglected for large times, except {\color{black} the Einstein tensor component}
\begin{eqnarray}
    G^2_0 \approx - \frac{e^x l_0}{a^3},
\end{eqnarray}
where $l_0 = l a$ is the present value of the rotation parameter. As $l_0 \ll \Omega_{m0}$ (the relative matter density, see below), this component is also negligible in the local universe ($x \ll 1$).
In the radiation era, when $l$ is constant, all the non-diagonal components of the {\color{black} Einstein tensor} fall slower than $a^{-4}$ and can be neglected in the limit $a \rightarrow 0$ as compared to the energy density and diagonal pressures, except
\begin{equation}
    G^0_2 \approx \frac{4l\epsilon\, e^{-x}}{3(1-l^2)},
\end{equation}
which is exponentially suppressed for $x \gg 0$.

\subsection{Radiation era}

As discussed previously, let us adopt for the radiation-dominated era the
ansatz $l =$ constant. The above equations turn out to be
\begin{equation} \label{radiation}
    \epsilon a^4 = - \left( 1 - \frac{3l^2}{4} \right) a^2 + 3\left( 1-l^2 \right) \dot{a}^2,
\end{equation}
\begin{equation} \label{p1}
    p_1 a^4 = p_2 a^4 = \left( \frac{l^2}{4} \right) a^2 + \left( 1-l^2 \right) \dot{a}^2 - 2\left( 1-l^2 \right)a \ddot{a},
\end{equation}
\begin{equation} \label{p3}
    p_3 a^2 = p_1 a^2 + 1 - \frac{l^2}{2}.
\end{equation}
Substituting into (\ref{radiation}) the conservation law for radiation, $\epsilon a^4 = a_0^2 =$ constant,
and considering the limit $a \rightarrow 0$, we obtain the solution
\begin{equation}
    a = b \eta = \sqrt{2bt},
\end{equation}
where $t$ is the cosmological time and
\begin{equation}
    b = \frac{a_0}{\sqrt{3(1-l^2)}}.
\end{equation}
The energy density is then given by
\begin{equation}
    \epsilon = \frac{3}{4t^2} (1-l^2).
\end{equation}
On the other hand, in the same limit $a \rightarrow 0$, Eqs. (\ref{p1}) and (\ref{p3}) give
\begin{equation}
    p_i = p = \frac{\epsilon}{3} \quad \quad \quad (i = 1, 2, 3).
\end{equation}
Therefore, the scale factor and energy density show the same time evolution as in the standard model, while the diagonal pressures present an isotropic equation of state.

\subsection{Matter era}

The matter dominated epoch is characterised by the conservation law $\epsilon a^3 = 2 a_1$,
with  $a_1$ constant. Adopting the ansatz $la=$ constant, taking the limit of large $a$ and keeping
only the dominant terms, the Einstein equations are reduced to
\begin{equation} \label{Friedmann}
    \epsilon a^3 = -a + \frac{3\dot{a}^2}{a},
\end{equation}
\begin{equation}
    p_1 a^3 = p_2 a^3 = \frac{\dot{a}^2}{a} - 2 \ddot{a},
\end{equation}
\begin{equation} \label{pressures}
    p_3 a^2 = p_1 a^2 + 1.
\end{equation}
From (\ref{Friedmann}) we obtain, apart from an integration constant, the solution
\begin{equation}
    a(\eta) = a_1 \left[ \cosh \left(\frac{\eta}{\sqrt{3}} \right) - 1 \right].
\end{equation}
With this solution, the remaining equations lead to
\begin{eqnarray}
    p_1 a^2 &=& p_2 a^2 = -\frac{1}{3}, \\
    p_3 a^2 &=& \frac{2}{3}.
\end{eqnarray}
We can see that {\color{black} pressures evolve slower than the matter density and, therefore, are subdominant except for late times, when a cosmological constant eventually dominates the expansion. In addition, the matter average pressure is zero at any time.}

\subsection{The energy content}

As discussed in \cite{saulo}, rotating solutions can be realised by a non-perfect fluid with heat conductivity and shear viscosity. The non-diagonal components of the energy-momentum tensor can be made negligible if the heat conductivity $\chi_0$ or the fluid temperature $T$ are small. On the other hand, the above pressure anisotropy gives rise to an anisotropy in the Hubble expansion given by ${h} = (2 \eta_0 a^2)^{-1}$, where $\eta_0$ is the shear viscosity. In the matter era the Hubble function decreases with $a^{3/2}$, and even slower when the cosmological constant enters into play. Therefore, for slowly varying viscosity, ${h}/H$ decreases with the expansion, which erases any strong signature of anisotropy.

In a local inertial frame whose origin is instantaneously comoving with the cosmic fluid, the energy-momentum tensor of an imperfect fluid can be written as \cite{Weinberg,Wheeler}
\begin{equation}
    T^{\mu \nu} = - p g^{\mu \nu} + (\epsilon + p) u^{\mu} u^{\nu} + \delta T^{\mu \nu},
\end{equation}
where $p$ is the average pressure, $u^{\mu}$ is the fluid $4$-velocity and $\delta T^{00} = 0$. The remaining corrections are given by
\begin{eqnarray} \label{Ti0}
    \delta T^{i0} &=& - \chi_0 (\partial^i T + T \dot{u}^i), \\ \label{Tij}
    \delta T^{ij} &=& - \eta_0 \left(\partial^j u^i + \partial^i u^j - \frac{2}{3} \partial_k u^k \delta^{ij} \right),
\end{eqnarray}
where, in this section, the overdot means derivative with respect to $t$.
In the case of metric (\ref{2}), the above equations reduce to
\begin{eqnarray}
    \delta T^{11} = \delta T^{22} &=& - \frac{2}{3} \eta_0 h, \\
    \delta T^{33} = \frac{4}{3} \eta_0 h,
\end{eqnarray}
where
\begin{equation}
    h \equiv \partial ^1 u^1 - \partial^3 u^3 = \partial^2 u^2 - \partial^3 u^3.
\end{equation}
In this way, we have
\begin{eqnarray}
    \delta T^{11} = \delta T^{22} = \delta T^{33} - 2 \eta_0 h,
\end{eqnarray}
which, compared to (\ref{pressures}), gives $2\eta_0 h a^2 = 1$, as anticipated above. On the other hand, the non-diagonal components of the energy-momentum tensor can be made negligible if the fluid temperature or the heat conductivity are small, as can be seen from (\ref{Ti0}), and by doing $u_i = H_i(t)\, x_i$ ($i=1,2,3$) in (\ref{Tij}). With these Hubble-Lema\^itre type laws we also obtain, from (\ref{Tij}),
\begin{equation}
H_1 = H_ 2 = H_3 - h,
\end{equation}
\begin{equation}
    H = (\partial_i u^i)/3 = \sum_i H_i/3.
\end{equation}

\section{Cosmological parameters}

In the previous equations describing the matter dominated era, we have neglected any rotation, as the rotation parameter decreases with the scale factor. However, even in the case that rotation is negligible today, it could be present at early times and our goal here is to verify its effect on the cosmological observables, in particular on the Hubble parameter\footnote{A Newtonian analysis of such an effect was previously presented in Ref.~\cite{newton}.}. Using the conservation laws $\epsilon a^3 =$ constant and $la=$ constant in (\ref{epsilon}), we can derive the Friedmann equation
\begin{eqnarray} \label{H}
    \epsilon = (3-l^2) H^2 - \frac{1-3l^2/4}{a^2},
\end{eqnarray}
where $H = \dot{a}/a^2$ is the Hubble parameter.

As CMB observations show no strong signature of spatial curvature, we can neglect the last term in a preliminary estimation. Assuming a non-zero rotation at the time of decoupling, we evaluate the relative difference in the Hubble function at that time as compared to the non-rotating case, for the same physical density. This leads to
\begin{equation} \label{deltaH}
    \frac{\delta H}{H} \approx 1 - \sqrt{1 - l_d^2/3}.
\end{equation}
On the other hand, it is possible to estimate a maximum value $l_d$ for the rotation parameter at the decoupling time by setting $l_{eq} = 1$ at the time of equality between matter and radiation, since $l$ is constant during the radiation era and for $l > 1$ we would have the presence of closed time-like curves \cite{Tiomno,Obukhov}. In this way, using $l_d a_d = l_{eq} a_{eq}$, 
we obtain $l_d \approx 0.3$. This would lead to a maximum correction of $2\%$ for the Hubble function at the decoupling time and to a corresponding higher value for the Hubble parameter when derived from CMB observations. 

Note, however, that the physics of decoupling depends not only on the density, but also on the expansion rate, and only a full fitting of the CMB spectrum of anisotropies can precisely constrain the cosmological parameters. Furthermore, the ansatz $l a =$ constant used in this estimation is only valid in the matter era. During the transition from radiation to matter dominated eras, $l$ falls slower and the condition $l_{eq} < 1$ would allow higher values for $l_d$ and $H_0$. More precisely, if we consider a cosmic fluid made by radiation and pressureless matter, the conservation of angular momentum $\epsilon a^4 l =$ constant leads to
\begin{equation} \label{suave}
    l = \frac{l_0 (\Omega_{R0} + \Omega_{m0})}{\Omega_{R0} + a \Omega_{m0}},
\end{equation}
where $\Omega_{m0}$ and $\Omega_{R0}$ are the present relative densities of matter and radiation, respectively. For $l_{eq} = 1$ this gives $l_d \approx 0.5$ and, from (\ref{deltaH}), a $4\%$ correction to the Hubble function at decoupling. Such a correction would not be enough to overcome the present gap between the values derived from CMB and Cepheid-based observations, but it could help to alleviate the tension, resulting in a value for $H_0$ in good agreement with, for instance, that derived with TRGB-based methods \cite{wendy}.

In order to perform a detailed analysis of CMB data that takes into account cosmic rotation, in a first approximation (i.e., considering the changes in the perturbation equations as higher order corrections) we add radiation and a positive cosmological constant to the left hand side of (\ref{H}), obtaining the Hubble function
\begin{eqnarray}
    \left(1-\frac{l^2}{3}\right) E(z)^2 = \Omega_{m0} (1+z)^3 + \Omega_{R0} (1 + z)^4 + \left( 1 - \frac{3l^2}{4} \right) \Omega_{k0} (1 + z)^2 + \Omega_{\Lambda},
\end{eqnarray}
where $z$ is the cosmological redshift, $E(z) = H(z)/H_0$, $\Omega_{(i)0}$ are the present relative densities of the energy components (including the spatial curvature), and $l = l_0 (1+z)$ for $z \ll z_{eq}$, $l_0$ being the rotation parameter at present, left as an additional free parameter. The above equation can thus be rearranged in the form
\begin{eqnarray}
    \left(1-\frac{l^2}{3}\right) E(z)^2 = \Omega_{m0} (1+z)^3 + \left( \Omega_{R0} - \frac{3l_0^2 \Omega_{k0}}{4} \right)  (1 + z)^4 + \Omega_{k0} (1 + z)^2 + \Omega_{\Lambda}.
\end{eqnarray}
Finally, let us note that, for $l_{eq} < 1$, 
we have $l_0 < 10^{-3}$. Since CMB observations restrict the {\color{black} $\Lambda$CDM} curvature parameter to $\Omega_{k0} < 10^{-3}$, we have $l_0^2\, \Omega_{k0} < 10^{-9}$. {\color{black} Even considering the possibility of a little higher spatial curvature, this term is} negligible as compared to $\Omega_{R0} \approx 8 \times 10^{-5}$. Therefore, the evolution equation for the Hubble function assumes the final form
\begin{eqnarray} \label{final}
    \left[ 1-\frac{l(z)^2}{3} \right] E(z)^2 = \Omega_{m0} (1+z)^3 + \Omega_{R0} (1 + z)^4 + \Omega_{k0} (1 + z)^2 + \Omega_{\Lambda},
\end{eqnarray}
with $l(z)$ given by (\ref{suave}),
\begin{equation} \label{suave2}
    l(z) = \frac{l_0 (\Omega_{R0} + \Omega_{m0})}{\Omega_{R0} + \Omega_{m0} (1+z)^{-1}}.
\end{equation}

\section{Results}

\begin{figure}[t]
\centerline{\includegraphics[scale=0.5]{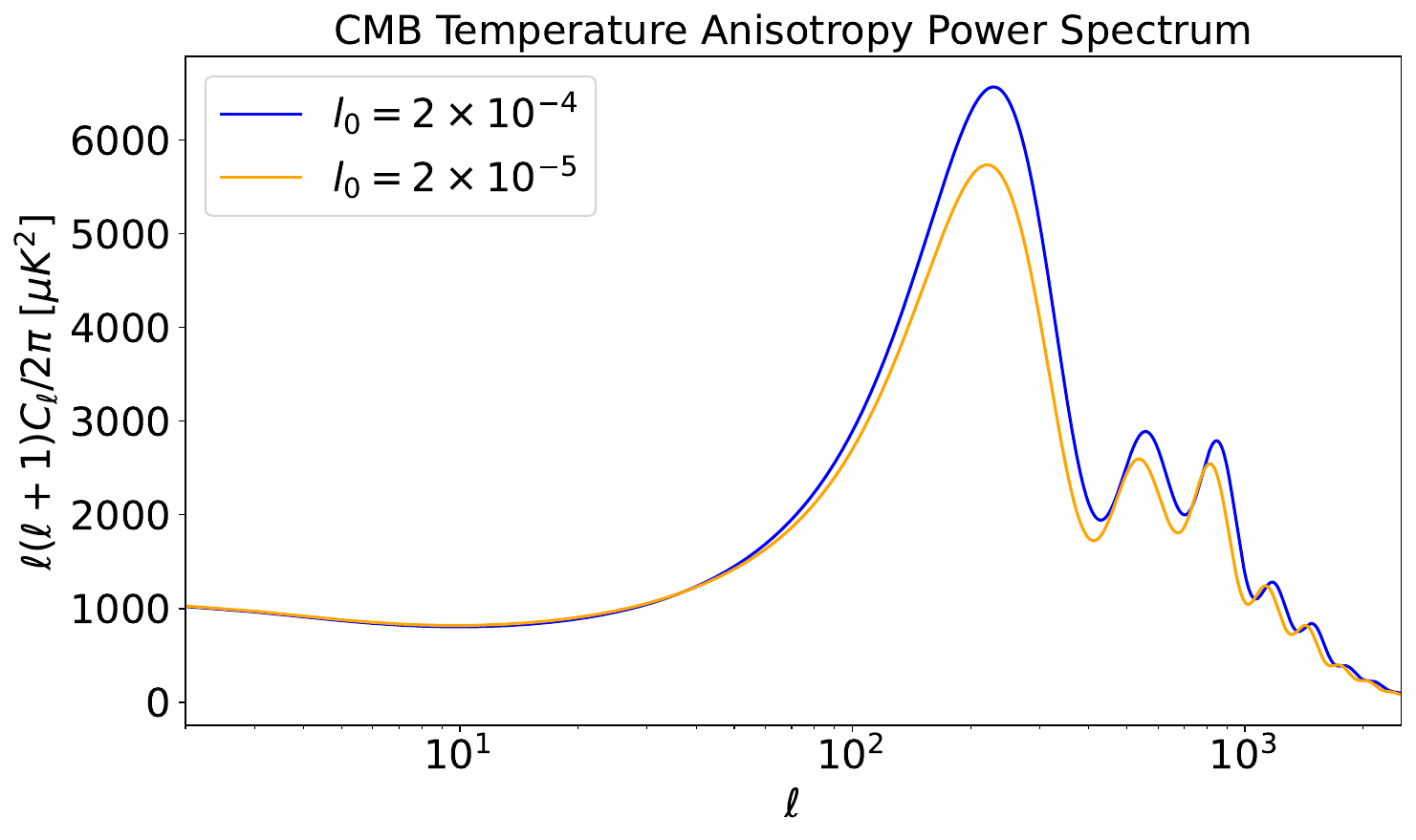}} 
\caption{CMB $TT$-spectra for $l_0 = 2 \times 10^{-5}$ (yellow) and $l_0 = 2 \times 10^{-4}$ (blue). All the other parameters were fixed in the standard model fiducial values.}
\label{Fig.1}
\end{figure}

The {\color{black} Monte Carlo Markov Chain (MCMC)} analysis was carried out using the COBAYA package \cite{Cobaya}, combined with a modified version of the CLASS Boltzmann code \cite{CLASS}, adapted to reflect the effects of rotation in the Newtonian gauge. 
We use the CMB dataset from the Planck 2018 release \cite{Planck:2019nip} to analyse early-time data. For late-time data, we combine the Pantheon Supernovae data \cite{Scolnic:2017caz} with BAO data from the 6dF Galaxy Survey (6dFGS) \cite{Beutler:2011hx}, the Sloan Digital Sky Survey (SDSS) Data Release 7 Main Galaxy Sample (DR7 MGS) \cite{Ross:2014qpa}, and the Sloan Digital Sky Survey (SDSS) Data Release 12 (DR12) ``Consensus" \cite{Alam:2016hwk} measurements. In this analysis using only late-time data, we treat the absolute magnitude of type Ia Supernovae as a free parameter within our cosmological fit, assuming a Gaussian prior on the direct local measurement of $H_0$ derived from Riess {\it et al.} (2021) \cite{Riess:2020fzl}, which is calibrated using Cepheid variables. 
The full analysis combining early- and late-time data, however, omits the Cepheid calibration of $H_0$ to avoid introducing tension between the probes, given the well-known discrepancy between local $H_0$ measurements and those predicted by CMB. This approach ensures a more consistent interpretation across the diverse datasets.
Hereafter, we will refer to these datasets as ``CMB", ``SNe+BAO" and ``CMB+SNe+BAO", respectively.
A Gaussian prior on the baryon density was imposed, $\Omega_{b0} h^2 = 0.02237 \pm 0.00070$ (here, $h \equiv H_0/100$ km/s-Mpc), in agreement with constraints from primordial nucleosynthesis \cite{Cooke:2017cwo}.

\textcolor{black}{Since the present-day rotation parameter can in principle span several orders of magnitude, we explore the range
\(0 \lesssim l_0 \leq 2.6 \times 10^{-4}\) using a logarithmic sampling. This choice ensures that the correction factor on the left-hand side of Eq.~(\ref{final}) remains positive and that \(l \leq 1\) at all redshifts, thereby preserving causality. For the remaining free parameters, we adopt the standard flat priors commonly used in \(\Lambda\)CDM analyses.}

For the scope of the present paper we only adapted the background modulus of CLASS, by redefining the physical densities as 
\begin{equation}
    \rho_{(i)} = \frac{\Omega_{(i)0} H_0^2 (1+z)^i}{1-\frac{l(z)^2}{3}},
\end{equation}
with $i = 0, 3, 4$ for the dark energy, pressureless matter and radiation components, respectively (the spatial curvature was fixed to zero\footnote{When we leave this parameter free, we obtain $\Omega_{k0} \approx 10^{-3}$ and compatible to zero at $1\sigma$.}). In this way, the Hubble function is written as $H(z)^2 = \sum_i \rho_{(i)}$, with constraint $H(z=0) = H_0$. As an illustration, Fig.~\ref{Fig.1} shows the $TT$ power spectra computed for two different values of the present-day rotation parameter, with all other parameters fixed to their fiducial values in the standard model.

\begin{figure}[t]
\centerline{\includegraphics[scale=0.3]{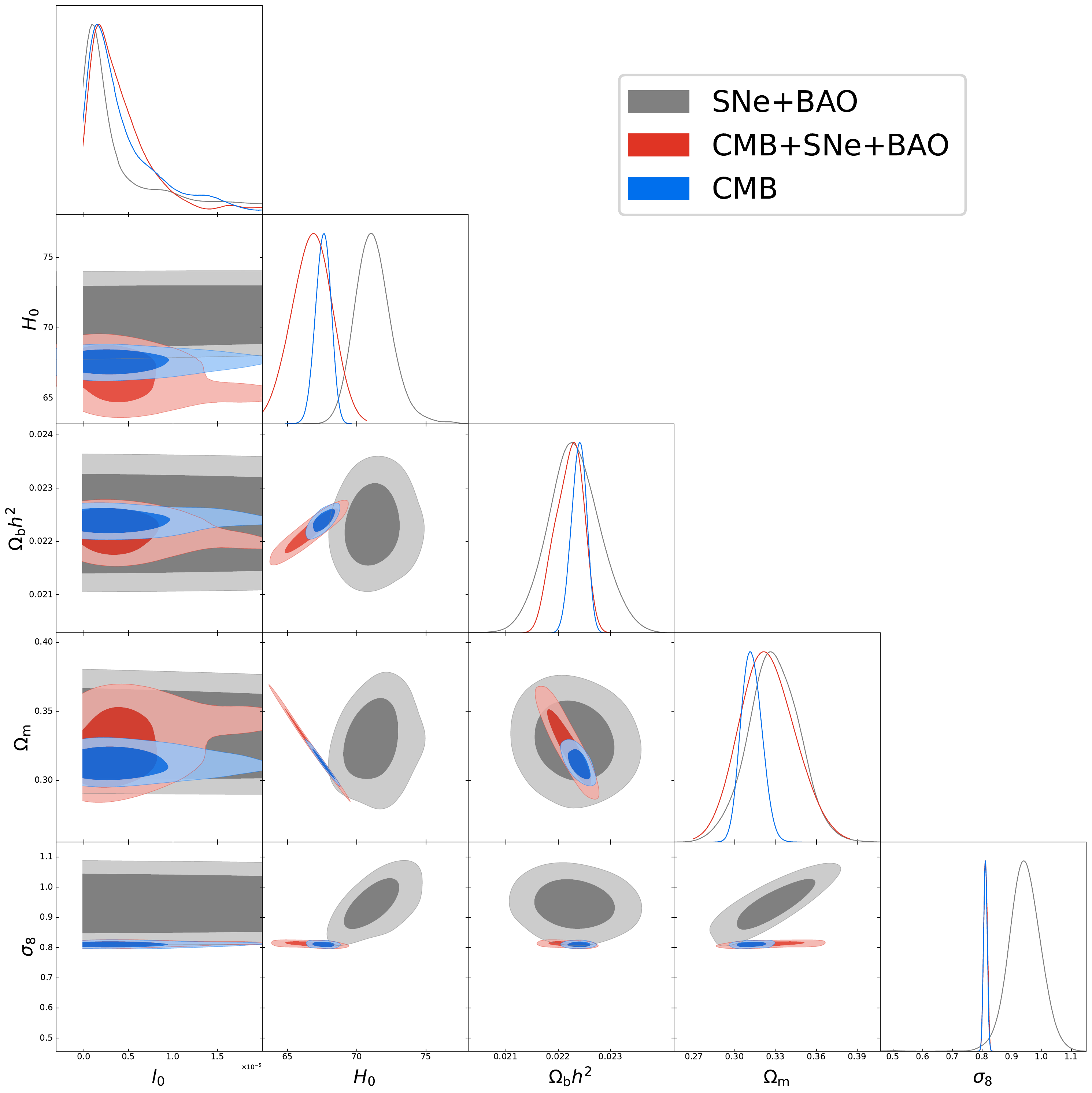}} 
\caption{Confidence regions for the rotating model using CMB Planck 2018 $TTTEEE$+$lowE$+lensing data \cite{Planck:2019nip}, Pantheon SNe \cite{Scolnic:2017caz} combined with BAO data \cite{Beutler:2011hx, Ross:2014qpa, Alam:2016hwk}, and a full analysis combining early- and -late time data. {\color{black} A SH0ES prior on $H_0$ \cite{Riess:2020fzl} has been imposed for the SNe+BAO analysis, but not for the full (CMB+SNe+BAO) analysis} A prior on the baryon density parameter was used in all the analysis \cite{Cooke:2017cwo}.}
\label{fig: triangle_plot}
\end{figure}

The results of the MCMC analysis are shown in Fig.~\ref{fig: triangle_plot}, which presents the 2D confidence regions for the main free parameters along with their posterior probability distributions. The corresponding mean values and 1$\sigma$ confidence intervals are listed in Table~\ref{tab:analysis}; for the present-day rotation parameter constrained by LSS data, only the upper limit is quoted, at $95\%$ confidence level.

From Tab.~\ref{tab:analysis}, we see that the CMB data constrain the rotation parameter to a value that is {\color{black} fully compatible with zero,}
therefore, we conclude that CMB does not favor a rotating model. Consequently, it is expected that all other parameters remain compatible with the standard model values -- as confirmed by the results, including the value of the Hubble constant.

The combined analysis of the CMB and SNe+BAO datasets doesn't significantly alter the central values of parameters constrained by CMB-only analyses. Instead, it widens the error bars, allowing for greater compatibility with higher $H_0$ values ($H_0<70.20$ km/s-Mpc at $99\%$ CL). This result is approximately $2.7 \sigma$ compatible with Riess's local measurement $H_0=73.04 \pm 1.04$ km/s-Mpc \cite{Riess:2020fzl}. On the other hand, the analysis based solely on the Pantheon SNe dataset \cite{Scolnic:2017caz}, combined with Barionic Acoustic Oscillation (BAO) measurements \cite{Beutler:2011hx, Ross:2014qpa, Alam:2016hwk}, allows for larger values of the present rotation parameter and yields a higher value of the Hubble constant. However, as in standard cosmological scenarios, a positive correlation between $H_0$ and $\sigma_8$ is observed, leading to a value of $\sigma_8 = 0.944 \pm 0.057$. This value shows a marginal tension of approximately $2.0\sigma$ with the DES-Y3 cosmic shear results \cite{DES:2021bvc} ($\sigma_8 = 0.802 \pm 0.020$) and is more compatible within about $1.3\sigma$ with the DESI galaxy survey constraint \cite{DESI:2024abc} ($\sigma_8 = 0.842 \pm 0.034$). {\color{black} Even taking into account that these observational constraints refer to the $\Lambda$CDM model, these discrepancies indicate, in view of the positive correlation found between $H_0$ and $\sigma_8$, a persistent tension between growth of structure measurements and local universe probes also within the model analysed here.

Finally, in the last two rows of Table~\ref{tab:analysis}, we show the $\Delta \chi^2$ and DIC values of the analysed model with respect to the reference $\Lambda$CDM model, considering the three datasets discussed in this work. 
As we can see, the $\chi^2$ values indicate a preference for the $\Lambda$CDM$ $ model (taken as the reference) with respect to the analysed model when using the {CMB} and {SNe+BAO} datasets. 
In contrast, the joint dataset {CMB+SNe+BAO} yields results statistically compatible with the $\Lambda$CDM model, showing no significant evidence in favor of either model. 
Regarding the DIC, it is computed as
\begin{equation}
\mathrm{DIC} = D(\bar{\theta}) + 2p_D = -2 \ln \mathcal{L}(\bar{\theta}) + 2p_D,
\end{equation}
where $D(\bar{\theta})$ is the deviance evaluated at the posterior mean of the parameters, and $p_D$ represents the effective number of parameters of the model. 
Therefore, the maximum-likelihood value is also weighted according to the model complexity, through the number of free parameters. 

A Jeffreys-type scale (e.g., $\Delta \mathrm{DIC} < 2$ indicates weak evidence, $2 < \Delta \mathrm{DIC} < 6$ moderate evidence, and $\Delta \mathrm{DIC} > 10$ strong evidence) is adopted for the interpretation of the results. 
In this framework, negative values of $\Delta\chi^2$ or $\Delta\mathrm{DIC}$ indicate that the reference $\Lambda$CDM model provides a better fit to the data. 
Accordingly, our results show a \textit{moderate statistical preference} for the $\Lambda$CDM model in the CMB and SNe+BAO analyses, while the combined dataset does not show any significant statistical preference between the two models.
}

\renewcommand{\arraystretch}{1.3}
\begin{table}[h!]
\centering
\begin{tabular}{lccc}
\hline
Parameter & CMB & SNe+BAO & CMB+SNe+BAO\\
\hline

$l_0$ & $(0.15^{+0.51}_{-0.15})\times 10^{-5}$ & $<1.66 \times 10^{-4}$ & $(0.16^{+0.42}_{-0.13}) \times 10^{-5}$ \\
$H_0$ & $67.57 \pm 0.53$ & $71.2 \pm 1.4$ & $66.7 \pm 1.2$ \\
$\Omega_b h^2$ & $0.02239 \pm 0.00015$ & $0.02230 \pm 0.00050$ & $0.02220 \pm 0.00026$ \\
$\Omega_\mathrm{m}$ & $0.3124 \pm 0.0071$ & $0.329 \pm 0.019$ & $0.3232 \pm 0.0177$ \\
$\sigma_8$ & $0.8190 \pm 0.0063$ & $0.944 \pm 0.057$ & $0.8112 \pm 0.0060$  \\
\hline
$\Delta \chi^2$ & $-4.4 $ & $ -3.4$ & $+0.5 $  \\
DIC & $-6.6 $ & $ -5.4 $ & $-1.5 $  \\
\hline
\end{tabular}
\caption{Mean values at $68\%$ C.L. using CMB Planck 2018 $TTTEEE$+$lowE$+lensing data \cite{Planck:2019nip}, Pantheon SNe \cite{Scolnic:2017caz} combined with BAO data \cite{Beutler:2011hx, Ross:2014qpa, Alam:2016hwk}, and a full analysis combining early- and late-time data. A prior on the baryon density parameter was used in all the analysis \cite{Cooke:2017cwo}. Upper limits are quoted at $95\%$ C.L.
{\color{black} The $\Delta \chi^2$ and DIC values shown in the last two rows refer to the comparison of the rotating model with the $\Lambda$CDM model, using the same dataset indicated in the column. A negative value of $\Delta\chi^2$ indicates that the reference model, $\Lambda$CDM, provides a better fit to the data. }}
\label{tab:analysis}
\end{table}


\section{Conclusion}

In this work, we explored the effects of cosmic rotation on the Hubble tension within a fully relativistic framework. We employed a rotating cosmological model sourced by a non-perfect fluid with anisotropic stress, originally proposed two decades ago in the context of a G\"odel-type spacetime. By modifying the background dynamics in the CLASS Boltzmann code, we were able to constrain the current cosmic rotation.

Our MCMC analysis based on Planck 2018 data places a stringent upper bound on the present-day rotation parameter, indicating a value {\color{black}compatible with} zero.
This limit implies that the inclusion of a small vorticity did not lead to any significant correction in the inferred value of the Hubble constant and, therefore, does not contribute toward alleviating the current tension between early- and late-time determinations of $H_0$, in spite of previous hopes based on a Newtonian study \cite{newton}.

Conversely, when considering late-time probes such as the Pantheon supernovae dataset combined with BAO measurements, larger values of the rotation parameter are allowed, yielding a higher inferred Hubble constant. However, as in standard cosmological scenarios, a positive correlation between $H_0$ and $\sigma_8$ emerges, with $\sigma_8 = 0.944 \pm 0.057$ in this case. This value remains in marginal tension at roughly $2.0\sigma$ with the DES-Y3 cosmic shear results \cite{DES:2021bvc} and is somewhat more compatible within approximately $1.3\sigma$ with the DESI galaxy survey constraint \cite{DESI:2024abc}. These discrepancies highlight a persistent tension between growth of structure measurements and local universe probes, which remains unresolved even within the rotating cosmology framework analysed here. 

Our comprehensive analysis, incorporating both CMB and SNe+BAO data, robustly affirms the parameter central values initially constrained by CMB-only observations. While maintaining consistency, this combined approach moderately loosens the precision of these parameter constraints, as reflected by an increase in their respective error bars.

Our analysis focuses solely on background modifications, preserving the standard treatment of linear perturbations. Extending this work to include rotational effects at the perturbative level is essential for a comprehensive understanding of large-scale structure formation within this framework. For example, the observed relation between angular momenta and masses of galaxies could be explained as a consequence of galaxy formation in an expanding and rotating universe \cite{Li}\footnote{\color{black} For other potential signatures of cosmic anisotropies, a recent review can be found in \cite{review2}.}.

In conclusion, our relativistic analysis shows that current CMB observations place strong constraints on the effects of cosmic rotation at the background level. Although a modest increase in the Hubble parameter could potentially reduce the discrepancy seen at the first acoustic peak in Fig.~\ref{Fig.1}, a full-spectrum fit imposes tight bounds on any deviation from the standard, non-rotating model -- as has generally been the case for most alternative cosmology attempts to resolve the Hubble tension. While a small rotational component may slightly increase the Hubble constant inferred from late-time data, it does not offer a significant or robust solution to the Hubble tension, highlighting the persistent difficulty faced by non-standard cosmologies in addressing this discrepancy.

{\color{black} Finally, we highlight two curious results that deserve a more physical interpretation. For a rotation parameter $l_0 \sim 10^{-3}$ and an observable universe radius of order $10^{26}$ m, the implied angular velocity $\omega_0 \sim 10^{-21}$ rad/s corresponds to an angular momentum $L \sim 10^{82}$ J·s. On one hand, this is the angular momentum of an extremal Kerr black hole with the mass of the observable universe, as also noted in \cite{newton}. On the other hand, it is $\Omega^3$ times the quantum of action \cite{saulo}, where $\Omega \sim 10^{40}$ is the Dirac factor involved in the so-called large number coincidences \cite{Dirac}. Although the various relations invoked in Dirac's hypothesis may arise from different physical mechanisms \cite{guillermo}, the presence of powers of a unique scaling factor is worthy of note. It is possible to argue that these two curious results are related: the horizon area of an extremal Kerr black hole, ${\cal A} = 8\pi L$, equals, in Planck units, the number of degrees of freedom encoded on the horizon. For a de Sitter universe, this number is $N \sim l_H^2/l_p^2 \sim 10^{120}$, where $l_H$ is the Hubble radius and $l_p$ the Planck length.

Even if such a rotation is not presently observed, note that it could be present in a pre-inflationary universe, being subsequently erased during an exponential expansion. From the conservation law $\epsilon a^4 l =$ constant we obtain, for a constant energy density, $l \propto a^{-4} \propto e^{-4Ht}$, where $H$ is the Hubble rate in the (approximately) de Sitter phase. After $N$ e-folds, an initial rotation parameter $l = 1$ would be suppressed by a factor $e^{-4N}$.}

\section*{Acknowledgements}

We are thankful to J.~S.~Alcaniz for useful suggestions. MB thanks support of the Istituto Nazionale di Fisica Nucleare (INFN), sezione di Napoli, iniziativa specifica QGSKY. DAC was supported with undergraduate grant from CNPq (Brazil). SC acknowledges support from CNPq with research grant 308518/2023-3. The authors thank the use of CLASS and COBAYA codes.

\end{document}